\RequirePackage{fix-cm}
\documentclass[smallextended]{svjour3}       
\usepackage[colorlinks,allcolors=blue]{hyperref}
\usepackage{amsbsy,amsmath}
\usepackage{amsfonts}
\usepackage{graphicx}
\usepackage{ulem}
\usepackage{breakcites}
\usepackage{lineno}

\title{Resonant perturbation of a family of young asteroids associated with (5026) Martes}
\titlerunning{Resonance perturbation of (5026) Martes}
\titlerunning{Resonance perturbation of (5026) Martes }

\author{Alexey Rosaev}
\authorrunning{A. Rosaev }

\institute{Alexey Rosaev \at Research and Educational Center ``Nonlinear Dynamics'', Yaroslavl State University, Yaroslavl, Russia\\ 
\email{hegem@mail.ru}   
}

\begin{document}

\date{Received: date / Accepted: date}

\maketitle



\begin{abstract}
The orbital dynamics of a very young asteroid pair (5026) Martes and 2005 WW113 is studied. We detect strong resonant perturbations of the larger member of the pair (5026) Martes by the 3:11 mean motion resonance with the Earth.  {The  second asteroid of the pair} (2005 WW113) has orbited far from the resonance and is not perturbed.  We provide a new estimation of the resonance structure and found that, under the planetary perturbations, a single resonance splits into a multiplet. 
The nominal position of the resonance is 2.37783 AU. However, the center of the corresponding chaotic zone is detected at 2.37754 AU. As a result, we noted a small but not negligible difference between the calculated and observed position of resonance. The multiplet structure of the 3:11 Earth resonance cannot explain this offset (the position of the main term of the multiplet is 2.377698 AU).

\end{abstract}



\section{Introduction}

The present paper is devoted to the investigation of the effect of the 3:11 mean motion resonance with the Earth on the motion of {the one of}the main belt asteroids. As a test particle, (5026) Martes is used. This asteroid is a member of a close and young asteroid pair with another small body 2005 WW113.

The close and young asteroid pairs are intensively studying recently (Pravec et al. (2019)). The study of such pair is more easy then old and dispersed asteroid families. Some examples of the resonance perturbations of young asteroid families and pairs are known, first of al, it is Datura family in 9:16 resonance with Mars (Nesvorny et al 2006). (Pravec et al. (2019)) note that pair (49791) 1999 XF31 and (436459) 2011 CL97 chaotic orbits may be explained by 15:8 mean motion resonance with Mars.  Duddy et al. (2012) pointed that pair (7343) Ockeghem and (154634) 2003 XX38 is in 2-1J-1M three body resonance.
 
As it was noted in paper Broz and Vokrouhlicky (2008),{ in the presence of the Yarkovsky effect} in resonance we have monotonic changes in eccentricity instead semimajor axis drift.  As a result, resonances can occur significant perturbations of orbits of some close asteroid pairs.

However the details of the resonance effect on evolution of orbits have not been completely understand. {In the previous article (see Table 1 in Rosaev (2022)), a comparison was made of determining the position of some resonances by different authors and by different methods. Variations in the position of the same resonance reach $\pm 0.0001$ AU.} Obviously, such accuracy is insufficient when studying compact asteroid pairs and families, where the distance between the members is approximately the same order.

 Previously we noted that {the 5026 Martes and 2005 WW113 pair} orbited close to the 3:11 mean motion resonance with the Earth Rosaev, Plavalova (2022). The main goals of this paper are to improve the estimation of 3:11E resonance position and to study the role of resonant perturbations in the dynamics of the (5026) Martes and 2005 WW113 pair. 

\begin{table*}[t]
\caption{The analytic proper elements and Lyapunov characteristic exponent (LCE) for the (5026) Martes and 2005 WW113 asteroid pair.Data of elements \textcolor{red}{downloads is} 30.10.2023}
\begin{center}
\begin{tabular}{l l r r r r r r} 
\hline 
 \multicolumn{2}{| c |}{Asteroid }  &   \multicolumn{5}{c |}{Proper elements}  \\
 \multicolumn{2}{| c |}{}&    \multicolumn{1}{c |}{e}   &  \multicolumn{1}{c |}{$LCE, 1/Myr$}  &  \multicolumn{1}{c |}{a, [AU]}  &  \multicolumn{1}{c |}{$s,''/yr$}  &  \multicolumn{1}{c |}{$g,''/yr$}  \\
  \hline \hline 
\multicolumn{1}{| c }{(5026)} & \multicolumn{1}{l |}{Martes}   & \multicolumn{1}{c |}{0.208072}  & \multicolumn{1}{c |} {40.57}  & \multicolumn{1}{c |} {2.37737}   &     \multicolumn{1}{c |}{-46.0335}    &     \multicolumn{1}{c |}{38.7909}\\
\multicolumn{1}{| c }{} & \multicolumn{1}{l |}{2005 WW113}  & \multicolumn{1}{c |}{0.207076}  & \multicolumn{1}{c |} {24.92}  & \multicolumn{1}{c |} {2.37697}   &     \multicolumn{1}{c |}{-45.9692}    &     \multicolumn{1}{c |}{38.7590}\\
 \hline 
\end{tabular}
\end{center}
\label{tabMartesProper}
\end{table*}

 \section{(5026) Martes and 2005 WW113 pair main facts}

 (5026) Martes and 2005 WW113 are listed in the paper by Vokrouhlicky and Nesvorny (2008) between  pairs with a low relative velocity. Later, (Pravec, Vokrouhlicky, 2009) noted that the pair is perturbed by irregular jumps over a weak mean motion resonance. But the resonance has still not been identified up for now. {Briefly, identification with 3:11 E resonance was mentioned in (Rosaev, 2023).}
 
Recently, the pair (5026) Martes and 2005 WW113 was studied by Pravec et al (2019), who found that this is a very young pair. Their backward integrations of heliocentric orbits suggest that these two asteroids separated about 18 kyr ago. 

The proper orbital elements of the pair obtained on the AstDys website Knezevic, Milani (2003) are given in the Table 1. {As can be seen from the value of the Lyapunov Characteristic Exponent (LCE) in Table 1, the orbit of 2005 WW113 is more stable than the orbit of 5026 Martes.}

As is it noted in (Pravec, Vokrouhlicky, (2009)), for so young pair it is possible to use osculating elements. We use model of initial orbital elements, according October 2023 (Table 2).

\begin{table*}[t]
\caption{\textcolor{blue}{The source of the data of the osculating elements. Epoch of osculation} 06.07.1998 (JD 2451000.5 TDB). \textcolor{blue}{Date of the elements downloads is 30.10.2023.}}
\begin{center}
\begin{tabular}{l l r r r r r r} 
\hline 
 \multicolumn{2}{| c |}{Asteroid }  &   \multicolumn{5}{c |}{Osculating elements}  \\
 \multicolumn{2}{| c |}{}&    \multicolumn{1}{c |}{e}   &  \multicolumn{1}{c |}{$i$, deg}  &  \multicolumn{1}{c |}{a [au]}  &  \multicolumn{1}{c |}{$\Omega, deg$}  &  \multicolumn{1}{c |}{$\omega, deg$}  \\
  \hline \hline 
\multicolumn{1}{| c }{(5026)} & \multicolumn{1}{l |}{Martes}   & \multicolumn{1}{c |}{0.243607}  & \multicolumn{1}{c |} {4.301214}  & \multicolumn{1}{c |} {2.3768484}   &     \multicolumn{1}{c |}{305.05020}    &     \multicolumn{1}{c |}{16.78112}\\
\multicolumn{1}{| c }{} & \multicolumn{1}{l |}{2005 WW113}  & \multicolumn{1}{c |}{0.242305}  & \multicolumn{1}{c |} {4.295131}  & \multicolumn{1}{c |} {2.3764550}   &     \multicolumn{1}{c |}{305.12615}    &     \multicolumn{1}{c |}{16.42209}\\
 \hline 
\end{tabular}
\end{center}
\label{Marosc_el}
\end{table*}

\begin{figure}
   \centering 
    \includegraphics[width=8.7cm]{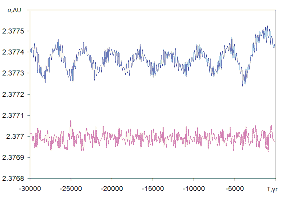}
     \caption{Evolution of semimajor axis (5026) Martes and 2005 WW113 nominal orbits.\textcolor{blue}{This shows the semimajor axis a in AU on the vertical axis and the time T in years on the horizontal axis. The evolution of (5026) Martes is shown in blue, 2005 WW113 in pink.} All planetary perturbations are included}
\label{cmp}
 \end{figure}

The pair 5026 Martes and 2005 WW113 is an example, when analytical proper elements have an advantage: the synthetic proper semimajor axis of both asteroid in the pair are equal, whereas in reality they differ in the observed time interval of 80 kyr: this is rightly reflected in the analytical proper elements (table 1, Fig.\ref{cmp}). Of course, over a long period of time, after a series of chaotic jumps, the  semimajor axis of 5026 Martes and 2005 WW113 may become very similar. 

Recently, four asteroids close to this pair have been discovered: 2010 TB155 (Novakovic et al. (2022)),  2011 RF40, 2022 QB59 and 2022 RM50 (Vokrouhlicky, et al, (2024)).

\section{Method and problem setting}

To study the dynamical evolution of this close asteroid pairs, the equations of the motion of the systems were numerically integrated 50 kyr into the past, using the N-body integrator Mercury (Chambers, (1999)) and the Everhart integration method (Everhart, (1985)).. To avoid short periodic perturbations, we employed a running average method for the results of numerical integration in a 1000 year window. To the nominal resonance position calculation, we use values of semimajor axis of planets, averaged over time of integration (30 thousands years in the past): 1.52368 AU for Mars, 5.20259 AU for Jupiter, 9.55490 AU for Saturn, 1.000001 AU for Earth. {The initial values of the semimajor axes of the planets were taken from the Horizon website. Thus, we provide a universal way to determine the position of the resonance.} 

To study interaction of the considered pair with the resonance and to determine {the position of the center of the instability zone corresponding to the resonance}  we apply the integration of orbits of the asteroid with the significant values of Yarkovsky effect ($A_2=1*10^{-13} AU/d^2$) and the different planetary perturbations.

Then, to estimate the accurate position of the resonance we integrate a several fictive orbits in vicinity of the resonance. {We use 5 fictive asteroids with the equal step 0.0001 AU in range of initial semimajor axis 2.37704-2.37744 AU and 4 fictive asteroids in range 2.3766-2.3769 AU. Other orbital elements are the same as for the 5026 Martes}
Additionally we have considered the effect of Ceres and Vesta on the 5026 Martes orbit and dependence of the result on the initial orbit variations.
   
\section{Resonance perturbations of pair (5026) Martes and 2005 WW113 }

The main asteroid of the pair - 5026 Martes shows strong resonant perturbations of the semimajor axis, which are {negligible} in the movement of 2005 WW113 (Fig.1). {As can be seen in  Figure 1, both asteroids have short periodic perturbations of their semimajor axis. However, semimajor axis of the 5026 Martes has an addition periodic perturbation with a large amplitude. In contrast, similar periodic perturbations on the semimajor axis of the 2005 WW113 have an amplitude similar to that of short periodic perturbations.} Moreover, the orbit of the 5026 Martes is unstable: when we use two systems of initial elements based on 1758 and on 3199 observation {(Dates of download from Horizon web site are 2019-Jul-21 and 2023-Oct-01 correspondingly)} we obtain different character of the semimajor axis variations. Therefore knowledge of resonant perturbations can play an important role in accurately estimation the age of the pair. For this reason, we will consider it in more details.

{The instability zone associated with this unidentified resonance is centered at 2.37750 AU and has a significant width of about 0.0002 AU in the semimajor axis, so the 5026 Martes could jump from one side of the exact resonance to the other. However, resonance has little or negligible effect on the second member of the pair.}
The resonance is not present in the atlas of Gallardo, (2014) and some other papers devoted to the three body resonances (Smirnov, Shevchenko, 2013) (the nearest 3-body resonance is 1 - 4J + 2S at 2.3967AU), so we can conclude that this is not a resonance of three bodies. The nearest low-order resonance with Jupiter is the 13:4 resonance (2.3714 AU); it is very distant. The closest resonance with Mars (20:39M at 2.3782088 AU, delta = 0.00075 AU) has too high an order.

\begin{figure}
   \centering 
    \includegraphics[width=8.7cm]{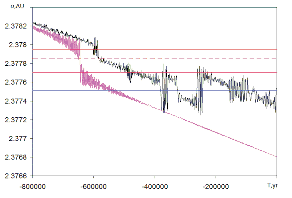}
     \caption{3:11E Resonance position in case of disturbances of only the Earth (pink) and all planets (black). {This shows the semimajor axis a in AU on the vertical axis and the time T in years on the horizontal axis.The dashed line marks the nominal position of resonance, the red straight lines marks the position of  the eccentricity and inclination resonances according the expression \ref{ae},\ref{ai}, the blue straight line marks the position of the resonance calculated by our method (expression \ref{are})}}
\label{RSc}
 \end{figure}

To the resonance identification, we have repeated our integration with a single planet as disturbing factor and with a very large value of the Yarkovsky drift to determine the actual position of the resonance. {The Yarkovsky effect causes the asteroid to approach the nearest resonance and cross it. However, under the influence of resonant perturbations, an asteroid can be temporarily captured into a libration orbit or deflected into an orbit before resonance. This process can be repeated several times which leads to a slowdown in the Yarkovsky drift.} As a result, we obtain that this is a single resonance of the mean motion with the Earth, because only in the case of a disturbance of the Earth do we have the destruction of the linear evolution of the semimajor axis of 5026 Martes under the action of the Yarkovsky drift. The position of the center of the resonance according to our numerical integration is about 2.377825 AU (Fig.\ref{RSc} ). 

To study the resonance in more detail, we conducted a repeated resonance search with independent data. {From Kepler's law, we get } that the 3:11 resonance with Earth a distance 2.377831 AU is the most likely main perturbation of the orbit of 5026 Martes. However, when we use integration with the same value of the Yarkovsky drift and perturbations of all planets, the position of the resonance is shifted to about 2.37755 AU according to Fig.\ref{RSc}.
The result differs slightly from our integration with only Earth perturbations (2.377825 AU), and from the paper by \cite{SD}  (2.377829 AU).

\begin{figure}
   \centering 
    \includegraphics[width=8.7cm]{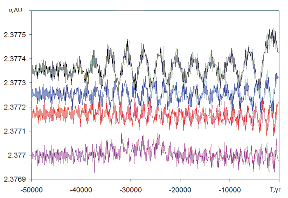}
     \caption{The results of the integration of fictive asteroids below the 3:11 resonance with Earth at a distance of 2.37745 AU (i.e. with $a<2.37745 AU$). {See text for more details about fictive asteroids. Here is the semimajor axis a in AU on the vertical axis and the time T in years on the horizontal axis.}}
\label{5027}
 \end{figure}
 \begin{figure}
    \centering 
     \includegraphics[width=8.7cm]{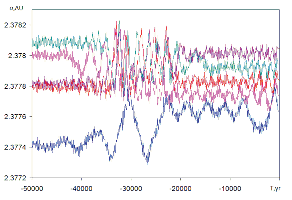}
      \caption{The results of the integration of fictive asteroids above the 3:11 resonance with Earth at a distance of 2.37745 AU (i.e. with $a>2.37745 AU$). See text for more details about fictive asteroids. Here is the semimajor axis a in AU on the vertical axis and the time T in years on the horizontal axis.}
 \label{5024}
  \end{figure}
  
  \section{Resonance position determination}
  \subsection{The preliminary estimation}
  It is interesting to estimate the position of the resonance by the frequency of the (5026) Martes orbit perturbations. The period of the perturbation (P) increases with the approach of exact resonance \cite{Sb}.
   
  To glean a clearer understanding of these dependencies, we applied the so-called Euclidian algorithm of resonance search for (see \cite{Sb} for a details). For a simple mean motion resonance is true:
  \begin{equation}
  j_1n_p+j_2n_a=0 ,
  \end{equation}
  
  where $ n_p, n_a$  are the mean motion of the planet and the asteroid respectively, $ j_1,j_2$ are integers. The most straightforward way to find $ j_1,j_2$ values is to use the decomposition of the mean motions ratio $n_p/n_a$ into continued fraction.
 
  \begin{equation}
  \frac{\overline{n_p}}{\overline{n_a}}=q_0+\frac{1}{q_1+\frac{1}{...+q_k}} .
  \end{equation}
   
 {Finally we have the approximation of the mean motion:}
     
  \begin{equation}
   \frac{\overline{n_p}}{\overline{n_a}}=\frac{q_0}{1}=\frac{q_0 q_1+1}{q_1}=...=\frac{j_2}{j_1}=\frac{c}{b}.
  \end{equation}
  
  It has long been established that the long periodic term of the perturbation (the so-called small divisors case) can have noticeable amplitude only when $j_1$ and $j_2$ have small absolute values. 
  In our case b=3, c=11 and the  small divisor is: $3n_p-11n_a$. Obviously, the corresponded period is: $P={2\pi}/(b n_p-cn_a)$. 

 We can substitute  $n_a=n_r+\delta_n$, {where $n_r$ is the mean motion corresponding
to the exact resonance and $\delta_n$  is the distance asteroid from the exact resonance (in mean motion). {For the exact resonance is true: $bn_p=cn_r$.} After that we can write} for the period of perturbation:

 \begin{equation}
 P=\frac{2\pi}{b n_p-c(n_r+\delta_n)}.
 \label{pr}
 \end{equation}
 
But for our goal is necessary to transform from the mean motion to the semimajor axis.
At the small $\delta_n $  it is possible to expand by power $\delta_n$:
 
   \begin{equation}
  \delta a=\left| a-a_r \right|\approx -\frac{2}{3}\delta_n a_r/n_r+... , 
  \label{delt}
   \end{equation}
   
   where $ a, a_r$  are the semimajor axis of the asteroid and the exact resonance respectively. 
 When the period P is known we have $\delta_n=2\pi/(cP)$. Substitute this value in the expression (\ref{delt}) for $\delta_a$  and resolve relative semimajor axis of resonance:
 
    \begin{equation}
     a_r\approx \frac{a}{1-2/3\delta_n/n_r}=\frac{a}{1- 4/(3cP/n_r)} .
     \label{are}
     \end{equation}  
  In our case: $n_r=3/11n_p=2\pi(3/11)$ .
  
 {Visually,} the period of the resonant perturbation at the current value of the (5026) Martes semimajor axis a= 2.37735 AU is about 4000 years. The resulting value of the exact resonant semimajor axis $a_r= 2.37748$ AU is in good agreement with the position of the center of the chaotic zone. This means that method of the resonance perturbations period can correctly estimate the nominal resonance position.  
  
     \subsection{The precision resonance position determination}
     
{However, the more accurate consideration is required.} Under the planetary perturbations, the 3:11 resonance acquires a multiplet structure. There are three main terms: eccentricity resonance:
  \begin{equation}
  a_{e}=(3/11+8\dot{\omega}/(11n_E))^{-2/3}a_E ,
  \label{ae}
  \end{equation}
  inclination resonance:
  \begin{equation}
  a_{i}=(3/11+8\dot{\Omega}/(11n_E))^{-2/3}a_E ,
  \label{ai}
  \end{equation}
  and resonance in the 3:11 E nominal position. {The upper line in Figure 2 where planets are used, also shows a change at a value of a=2.377825 AU.}
  
 {Here $\dot \omega $ is the rate of the perihelion longitude change and $\dot \Omega $  is the rate of the node longitude change. For more information about eccentricity resonance and inclination resonance see the book by \cite{MD}} 
  
  But the eccentricity and inclination resonances are shifted relative to their calculated position. {If we substitute $\dot \omega=g$, $\dot \Omega=s $ (see table 2), we obtain} $a_e= 2.377698 AU $ and $a_i= 2.377975 AU $ respectively.
  The eccentricity resonance, which is most important for us, formed a instability zone at about 2.37755 AU. 
  
  We study the boundary between the terms and their detailed positions by integration of a several fictive asteroids (Fig.\ref{5027}-\ref{5024}). The last figure is the most interesting. {In the right central part of it (interval 0-20 thousand years ago) we see an increase in frequency with an increase in the semimajor axis. However, starting from the semimajor axis about a=2.37785 AU, another periodic disturbance appears.}  Consequently, we can detect two terms of the resonance with semimajor axes about a=2.3775 AU and a=2.3779 AU. The last value coincides with the nominal position of the 3:11 resonance, but this perturbation is very weak. The possible boundary between these two terms is about 2.3778 AU. In the epoch between 20 and 35 kyr ago we see {variations of semimajor axis of all clones with large amplitude}. This instability is corresponds to the upper term at a=2.3779 AU. 
  
  It is possible to point out the lower boundary when the resonance effect becomes negligible, about 2.3770 AU, which is just close to the orbit of 2005 WW113. 
  
  Solving equation (\ref{delt}) above relative $\delta_n$ and substitute in (\ref{pr}) we have:
 \begin{equation}
 P= \frac{2\pi}{c\delta_n}=\frac{4\pi a_r}{3c\delta_a n_r}.    
 \end{equation}
  After taking into account $bn_p=cn_r$,  for the frequency we have:
   \begin{equation}
 f=\frac{1}{P} \approx\frac{3bn_p}{4\pi a_r}\left| a-a_r \right|.
  \end{equation} .
  The frequency of resonance perturbations is proportional to the distance from exact resonance in semimajor axis.
  
On base of the integration of fictitious asteroids (Fig.\ref{5027}-\ref{5024}), we provide an accurate determination of the position of the resonance. {The frequency value was searched by the criterion of minimum standard errors calculated by the expression:}
\begin{equation}
    S=\sqrt{\frac{1}{N-2}\sum_{i=1}^{N}\left(a_{i}-a_{i_{appr}} \right)^{2}},
\end{equation}
where N is the number of output points in integration, $a_i$ is the value of the semimajor axis during integration, and $a_{i_{app}}$ is the corresponding approximate value.

\begin{figure}
      \centering 
       \includegraphics[width=8.7cm]{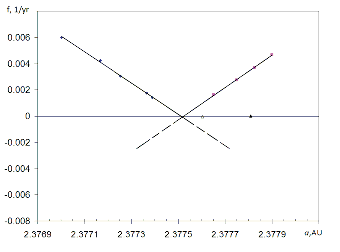}
        \caption{The results of the integration of fictive asteroids above and below the resonance. {This shows the frequency \textit{f} (see text) in 1/yr on the vertical axis and the semimajor axis \textit{a} in AU  on the horizontal axis.} The nominal position of the 3:11E resonance is marked by a filled triangle, the position of the eccentricity multiplet term is marked by a free triangle.{The blue squares represent the fictive orbits, corresponding to Figure (3), the pink squares correspond to  the fictive orbits in Figure (4)}}
   \label{SL0}
    \end{figure}

The dependences of frequency of the semimajor axis perturbations, above and below the instability zone, are accurately approximated by straight lines each with its own coefficients (Fig.\ref{SL0}):
\begin{equation}
    f_b=-11.938a+28.383,
\end{equation}
\begin{equation}
    f_a=12.335a-29.327.
\end{equation}
\textcolor{blue}{Here $f_a, f_b $ are the frequencies above and below the resonance.}

The intersection of these two lines corresponds to an infinite period of perturbations, i.e. motion along the separatrix. The equations of the approximating straight lines give for the zero frequency the values 2.3775339 AU and 2.3775436 AU respectively, and for the intersection point the value a= 2.3775388 AU. Therefore we can associate this to the exact resonance position at 2.37754 AU. Of course, the real resonance has a finite width but we note the displacement between the obtained value and both the nominal and multiplet position. The explanation of this may be the subject of further study. The main result of this paper is the quantitative determination of this offset. 

However one important note is evident: the offset is not associated with any non-gravitational force and with any process of migration: it is present in the pure gravity problem.  Of course, the numerical value will depend on the initial conditions but the principal result will remain unchanged.

 \section{The effect of eccentricity}

The next important questions are to explain the instability close to -27 kyr. The eccentricity of Martes has maximal value close to 27 kyr ago (Fig.6). As it is known, the {width of libration zone increases with eccentricity (see figure 8.7. in the book of \cite{MD}). It means that orbit with large eccentricity can become close to resonance and a temporary capture/escape in resonance can take place.}. 
This process can initialize chaotic behavior of the Martes semimajor axis about 27 thousand years ago. 

We can outline that resonance is very narrow, not more than with $1*10^{-4}$ AU widths, and has high order. But we may guess that this resonance is related to the process of forming considered pair. 

After the Ceres and Vesta perturbations taking into account, we obtain similar oscillations of semimajor axis with rather smaller period. This means that perturbations of large asteroids slightly reduce the mean semimajor axis of Martes but do not change the resonance position. Note that the disturbances of Ceres and Vesta do not affect the second asteroid in the pair in any way.

   \begin{figure}
      \centering 
       \includegraphics[width=8.7cm]{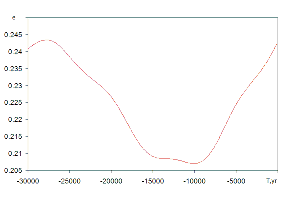}
        \caption{Eccentricity evolution of (5026) Martes. {Here is the eccentricity e on the vertical axis and the time T in years on the horizontal axis
}}
   \label{Ex0}
    \end{figure}
\section{Result of new members orbits integration}

Recently, four asteroids close to this pair have been discovered: 2010 TB155 (\cite{No}),  2011 RF40, 2022 QB59 and 2022 RM50 (\cite{Ve}). Consequently, the group associated with 5026 Martes becomes a very young family. The discovery is very important for understanding the origin of this cluster because the direct separation of 2005 WW113 from 5026 Martes requires notably large relative velocity or unrealistic values of the Yarkovsky effect.

Here we study these new members with an emphasis on their resonant perturbations. First, we integrate the orbits of the new members with the perturbations of the large planets only. 

Note that the three of new members (2011 RF40, 2022 QB59 and 2022 RM50) orbited closer to the 3:11E resonance as well as 5026 Martes. Therefore their separation is easier than 2005 WW113 and 2010 TB155. Two new members (2022 QB59 and 2022 RM50) are moving in an almost identical orbit. The result is shown in fig 7.

The minimum distance between 2022 QB59 and 2022 RM50 is about 4500 km in the epoch 17.003 thousand years ago. Despite the intersection of the resonance about 3, 6, 16, 21 thousand years ago, the distance between them does not exceed 0.28 AU during the entire considered interval (Fig.8). This means that 2011 RF40, 2022 QB59 and 2022 RM50 orbited in a very stable region of phase space which is interesting in itself. However, the closest encounter with 5026 Martes occurred an about 17.45 kyr ago.
\begin{figure}
      \centering 
       \includegraphics[width=8.7cm]{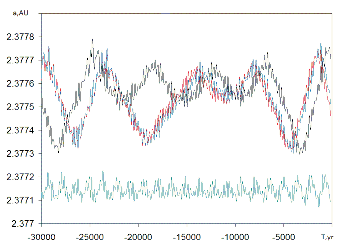}
        \caption{Evolution of semimajor axes of new members of Martes group. Only planet perturbations.}
   \label{nmb}
    \end{figure}
\begin{figure}
      \centering 
       \includegraphics[width=8.7cm]{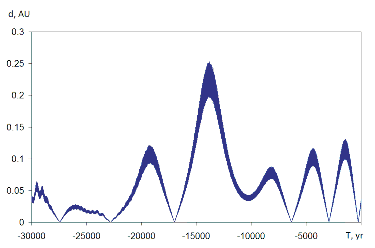}
        \caption{Distance between 2022 QB59 and 2022 RM50. }
   \label{Db}
    \end{figure}

 \begin{figure}
      \centering 
       \includegraphics[width=8.7cm]{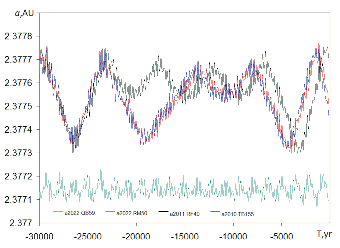}
        \caption{Evolution of semimajor axes of new members of Martes group.Planets and Ceres and Vesta perturbations.}
   \label{nmb}
    \end{figure}
    
This conclusion is confirmed by our integration with Ceres and Vesta effect. The orbits of  2022 QB59 and 2022 RM50 remain unchanged, while the orbits of 2011 RF40 and 5026 Martes slightly change the mean semimajor axis (Fig.9). Moreover, the orbit of 2011 RF40 in the time interval between 31 and 24 thousand years ago becomes the same as the orbits of 2022 QB59 and 2022 RM50.
 
    The minimum distance between 2022 QB59 and 2022 RM50 is about 1720 km at 17.075 kyr. However, the closest encounter with Martes takes place an about 15.45 kyr ago for this case.

    The long term integration confirms a very stable character of orbits of 2022 QB59 and 2022 RM50 up to 800 kyr in the past (Fig.10).

    \begin{figure}
      \centering 
       \includegraphics[width=8.7cm]{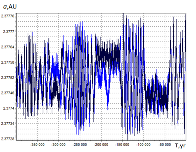}
        \caption{The long-term evolution of the semimajor axis of 2022 QB59 (black) and 2022 RM50 (blue)}
   \label{Ex0}
    \end{figure}
\section{Relation with the problem of the pair origin}

Integrations of the Martes orbit with different initial conditions show that it is very unstable. The resonant instability of the orbit of 5026 Martes leads to an uncertainty of the orbital position by about 180 degree in longitude in the time interval under consideration ({30 thousands years}, \cite{Ro}). Obviously, this has a big impact on the assessment of the couple's age.

 Taking into account the perturbations of Ceres and Vesta, we confirm this conclusion. The small variations in the initial orbit or perturbations lead to noticeable changes in the Martes longitude during the forming encounter. In this context, the obtained relationship between the semimajor axis and the frequency of perturbations can help to better determine the position of Martes in the epoch of breakup. 


\section{Discussion and conclusions}

The orbital dynamics of a very young asteroid pair (5026) Martes and 2005 WW113 is studied. We detect strong resonant perturbations of the larger member of the pair (5026) Martes by the 3:11 mean motion resonance with the Earth.  The unbounded secondary (2005 WW113) has moved far from the resonance and is not perturbed.  We provide a new estimation of the resonance structure and found that, under the planetary perturbations, a single resonance splits into a multiplet.
The nominal position of the resonance is 2.37783 AU. However, the center of the corresponding chaotic zone is detected at 2.37754 AU. As a result, we noted a small but not negligible difference between the calculated and observed position of resonance. The multiplet structure of the 3:11 Earth resonance cannot explain this offset (the position of the main term of the multiplet is 2.377698 AU). 

The resonance is displaced relative nominal position and the offset cannot be explained by the planetary perturbations and non-gravitational force. This resonance is necessary to account during the study of the pair origin. 

\section{Declaration}
Conflict of interest The author declares that he has no conflict of interest



\bibliographystyle{spphys}       

\end{document}